\documentclass{sigchi-ext}
\usepackage[T1]{fontenc}
\usepackage{textcomp}
\usepackage[scaled=.92]{helvet} 
\usepackage{graphics} 
\usepackage{balance}  
\usepackage{booktabs} 
\usepackage{ccicons}  
\usepackage{ragged2e} 
\usepackage{todonotes}
\usepackage{url}
\usepackage{txfonts}
\usepackage{mathptmx}
\usepackage{color}
\usepackage{csquotes}
\usepackage{microtype}        
\usepackage[all]{hypcap}    

\def\plaintitle{Blockchain for Trustful Collaborations between Immigrants and Governments} 
\def\emptyauthor{}
\def\plainkeywords{Authors' choice; of terms; separated; by
  semicolons; include commas, within terms only; required.}

\title{Blockchain for Trustful Collaborations between Immigrants and Governments}

\numberofauthors{6}
\author{%
  \alignauthor{%
    \textbf{Chun-Wei Chiang}\\
     \affaddr{HCI Lab} \\
    \affaddr{West Virginia University}
    \email{cc0051@mix.wvu.edu} }\\\\
  \alignauthor{%
    \textbf{Eber Betanzos}\\
    \affaddr{Secretaria de la Funcion Publica}\\
    \email{ebetanzost@presidencia.gob.mx} }\\\\
   \alignauthor{%
    \textbf{Saiph Savage}\\
    \affaddr{HCI Lab}\\    
    \affaddr{Universidad Nacional Autonoma de Mexico (UNAM)}
    \email{saiphcita@gmail.com} }\\\\
    }

\definecolor{linkColor}{RGB}{6,125,233}
\hypersetup{%
  pdftitle={\plaintitle},
  pdfauthor={\emptyauthor},
  pdfkeywords={\plainkeywords},
  bookmarksnumbered,
  pdfstartview={FitH},
  colorlinks,
  citecolor=black,
  filecolor=black,
  linkcolor=black,
  urlcolor=linkColor,
  breaklinks=true,
}

\newcommand{\sys}{ChainGov }
\begin{document}


\CopyrightYear{2018} 
\setcopyright{rightsretained} 
\conferenceinfo{CHI'18 Extended Abstracts}{April 21--26, 2018, Montreal, QC, Canada}\isbn{978-1-4503-5621-3/18/04}
\doi{https://doi.org/10.1145/3170427.3188660}
\copyrightinfo{\acmcopyright}

\maketitle

\RaggedRight{} 

\begin{abstract}

Immigrants usually are pro-social towards their hometowns and try to improve them. However, the lack of trust in their government can drive immigrants to work individually. As a result, their pro-social activities are usually limited in impact and scope. 
This paper studies the interface factors that ease collaborations between immigrants and their home governments. We specifically focus on Mexican immigrants in the US who want to improve their rural communities. We identify that for Mexican immigrants having clear workflows of how their money flows and a sense of control over this workflow is important for  collaborating with their government. Based on these findings, we create a blockchain based system for building trust between governments and immigrants. 
We finish by discussing design implications of our work and future directions.
\end{abstract}



\section{Introduction}

In 2013 there were at least 13 million Mexican immigrants in the US \cite{ratha2016migration}. The amount of money that these immigrants send back home is the fourth largest in the world, amounting to around 25.2 billion US dollars  \cite{ratha2016migration}. 
Remittances not only assist immigrants' families but can also facilitate community development \cite{orozco2004mexican}. For decades, Mexican immigrants constructed various projects to benefit their native communities with their knowledge \cite{kuznetsov2006diaspora} or wealth \cite{de2005redefining}. 
The donations from Mexican immigrants are usually very meaningful to their hometowns. Sometimes the donation can be even 7 times the budget assigned to the town's local government \cite{kuznetsov2006diaspora}. 

 \begin{marginfigure}[4pc]
   \begin{minipage}{\marginparwidth}
     \centering
     \includegraphics[width=0.9\marginparwidth]{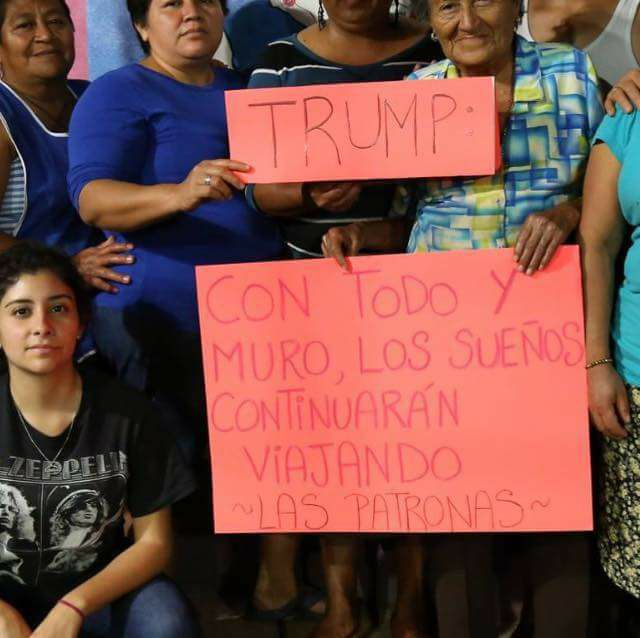}
    
     \caption{Example of a citizen group, ``Las Patronas'', working in the US and Mexico to help rural migrants and the families they leave behind. Here, the women engage in activism. They share signs to support migrants:\emph{``Trump: even with the wall, the dreams will continue traveling.''} However, these groups usually lack official support from governments and therefore have more limited impact. Photo source: Facebook.}~\label{fig:patronas}
  \end{minipage}
 \end{marginfigure}

However, despite their good intentions, most of these efforts usually have little impact and are executed on a very small scale \cite{kuznetsov2006diaspora}. One of the reasons is that they generally lack institutional support: the government is not behind their efforts. The main issue is that a majority of Mexican immigrants distrust the Mexican government and institutions \cite{kuznetsov2006diaspora,morris2009political}. Therefore, they lack the willingness to collaborate with them and prefer to supervise and execute the projects by themselves \cite{kuznetsov2006diaspora}. We believe that a way to increase the impact that immigrants have over their hometowns is by creating alternative tools that facilitate their political participation \cite{friedmann1992empowerment}. We specifically study tools that build trust between immigrants and their governments. 

In this late-breaking work, we investigate the interface factors, especially those that are known to be important in Latin America \cite{moreno2016breaking}, that can motivate or hinder Mexican immigrants to contribute their finances and collaborate with the Mexican government to aid their hometowns. We then use the findings of our study to design \sys a system that helps immigrants, NGOs, and local governments to cooperate with each other for community prosperity, by: giving citizens more agency over the finances they donate; fighting corruption; and enhancing fiscal transparency in community development projects. We finish by discussing design implication of our research. 


\section{Investigating Interface Factors for Facilitating Government-Immigrant Collaborations}

Here we investigate how technology could ease collaborations between immigrants and governments. We consider that immigrants have access to mobile phones
, and could potentially use these devices to collaborate with their governments to help their hometowns.  Previous work showed that individuals trust can be affected by the user interface \cite{zheng2016context}. We believe that such collaborations could especially be enabled with mobile interfaces that facilitated trust building. 
We thus analyze the interface factors of mobile money, i.e., interfaces that allow immigrants to transfer their wealth via mobile devices to their hometown. We especially investigate the mobile money interface factors that facilitate trust-building. 

\subsection{{\bf Evaluation}}
\begin{marginfigure}[0pc]
  \centering
  \includegraphics[width=.8\columnwidth,height=8cm,keepaspectratio]{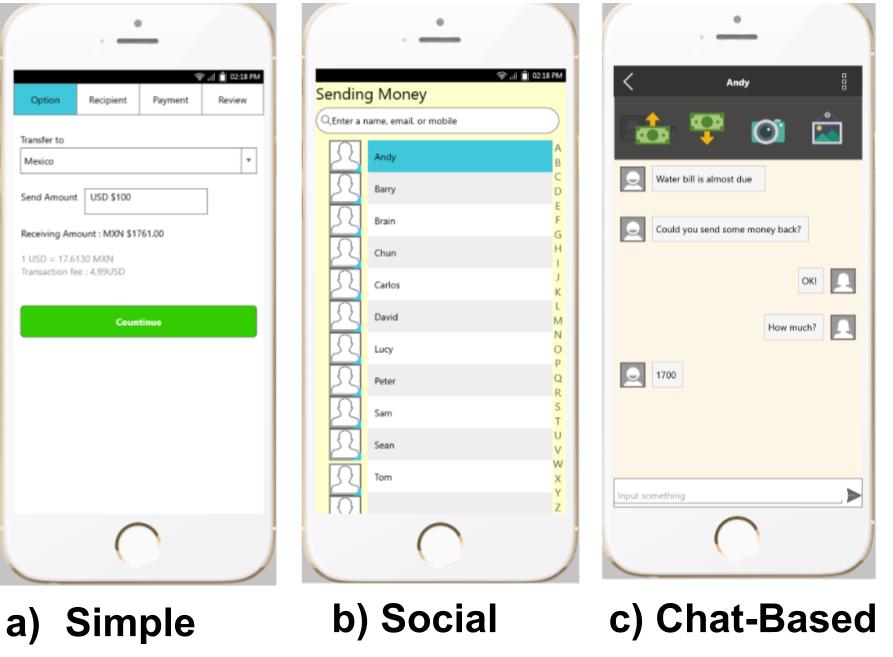}
  \caption{Overview of the interfaces we presented to participants based on the different types of mobile money interfaces that prior work had identified: a) Simple Interface Model; b) Social Interface Model; c) Chat-based Interface Model.}~\label{fig:model}
\end{marginfigure}

We investigate the perceptions that Mexican immigrants have about the 3 main types of mobile money interfaces identified by prior work (see Figure~\ref{fig:model}) \cite{chiang2017understanding}. We especially study how each interface facilitates (or hinders) trust building. 
The interfaces we study are: ``Simple'' interface model which is tailored to just do the task of transferring money to the destination (nothing else); the ``Social'' interface model presents a simple interface model and also a view where people can visualize the money transfers of their friends or contacts; the ``Chat-Based'' interface model where participants can send money transfers via chat. 

\subsubsection{Participants}
To recruit participants we conducted a ``street-intercept survey'' during large-scale events involving Mexican immigrants in the US. We recruited a total of 88 Mexican immigrants. Their age ranged between 18 and 40 year olds (M = 24.13, SD = 4.80, Median = 22.92). 30\% of the participants were female and 70\% were male. 39\% of our participants had more than 6 years of experience in using mobile phones, 46\% had between 4-6 years of experience in using mobile phones, and 15\% reported they had less than 3 years using mobile devices. Our participants had varying degrees of experience with mobile money and international remittance services (especially services for sending money back to Mexico). 

\begin{marginfigure}[2pc]
  \begin{minipage}{\marginparwidth}
    \centering
    \includegraphics[width=1.0\marginparwidth]{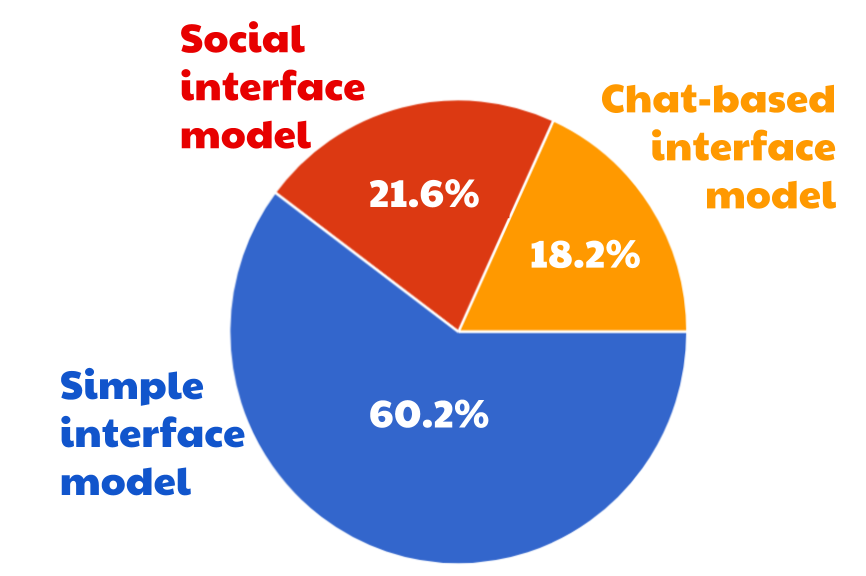}
    \caption{Overview of the percentage of participants who preferred a particular type of mobile money interface because of the trust it inspired. Most participants considered the  ``Simple Interface Model'' to be the most trustful.}~\label{fig:modelresult}
  \end{minipage}
\end{marginfigure}
\subsection{Study}
Our study had two main parts: (1) investigate the experiences of participants with mobile money interfaces,  transferring money to Mexico and collaborating with the Mexican government; and (2) have participants use each mobile money interfaces to then question them more deeply about how they felt the interface facilitated/hindered trust building and sending money to Mexico.

\subsubsection{1. Immigrants' Mobile Money Experiences.} We first collected information about participants' background knowledge on mobile money, such as the frequency of using international remittance service, and mobile financial applications. The survey also questioned participants about their habits for transferring money and how much they trusted different money transfer services. 


\subsubsection{2. Immigrants' Interactions with Mobile Money.} Here we had participants use the 3 different mobile money interfaces. Afterwards, we asked them to compare the interfaces and evaluate which model gave them more trust for sending money back home to collaborate with their native government. We counterbalanced the order in which we showcased each interface to participants. We also asked participants several sequential questions about their thoughts on specific features on the 3 different mobile money interfaces. We especially questioned participants' perceptions of how each feature might facilitate trust building. We choose these features (i.e., interface factors) because prior work identified they were important in user adoption of money applications \cite{zheng2016context}.

\section{Results}

Overall, 43 \% of our participants transferred money to Mexico through online banking services. 27\% preferred to transfer money through brick and mortar financial services (i.e., offline services). However, all of these individuals did have experience operating online money transfer services. 30\%  preferred to not use online banking services, but instead used online services provided by other institutions, e.g., bitcoin. 

Overall it seemed that participants had more trust issues with non-traditional institutions than well-established organizations. For instance, participants expressed they had confidence in bank employees (mode = 5, median = 4). But participants, in general, did not trust employees from non-traditional financial institutions (mode = 3, median = 3), such as from the Western Union or PayPal. 

All of our participants had used previously the 3 different types of mobile money interfaces. However, the majority felt the Simple Interface Model was the most trustful. In specific, 60\% felt the Simple Interface inspired the most trust, 21\% the Social, and 18.2\% the Chat-Based Interface. We did not observe a particular age group to prefer one interface over the other (see Figure~\ref{fig:modelresult}).  

\begin{marginfigure}[-8pc]
  \begin{minipage}{\marginparwidth}
    \centering
    \includegraphics[width=0.9\marginparwidth]{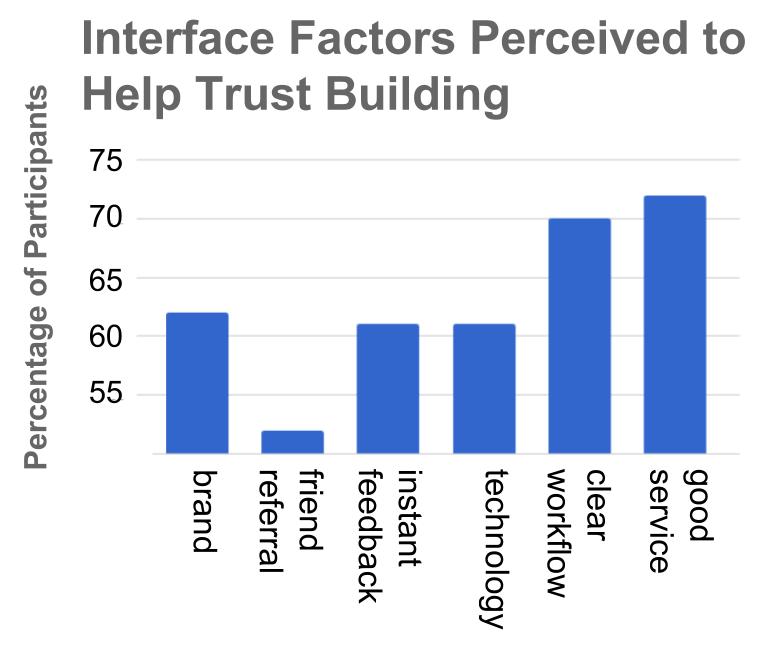}
    
    \caption{Overview of the percentage of participants who consider that certain mobile money interface features were important for building trust and facilitating collaborations with the government. Clear workflow and good service were the most important features across mobile money interfaces.}~\label{fig:moniGraph}
  \end{minipage}
\end{marginfigure}

We also identified that participants did not trust technology to interact with their finances. The majority of our participants (42\%) distrusted online financial services, regardless of whether the services were offered by banks (mode = 4, median = 4) or other financial institutions (mode = 3, median = 3). The preferred mode of interaction for accessing their finances (66\%) was with human agents who could ensure them that everything was in order and who could rapidly respond to all their questions, especially of the status of their money. For Mexican immigrants, it seemed important to have a sense of control and understand how their finances were moving. This point also appeared to influence their willingness to collaborate with the government. Participants appeared to be open for collaborations if they had an assurance of how their finances were used at all times. Participants seemed to prefer the Simple Interface because it allowed them to have more control over how their finances moved. 
Our study also revealed (see Fig.\ref{fig:moniGraph}) that across mobile money interfaces, ``good service'' and ``clear workflow'' (i.e., being able to visualize how their money moved) were for participants the most important interface features for trust building and cooperating with the government. Overall, our results suggest that by providing transparent visualizations and agency, we can help Mexican immigrants to trust and collaborate with their home government more through mobile money applications.

\section{Designing Systems for Immigrant-Government Collaborations}

From our study, we identified that Mexican immigrants had trust issues with technology and institutions (even more so with non-traditional institutions). Mexican immigrants thus seemed to value transparency in their mobile money interfaces. They especially wanted to clearly visualize the flow of their finances (i.e., how their money moved). We use our findings as a design probe to create systems that lead to trust building and ease collaborations between immigrants and governments.

\begin{marginfigure}[0pc]
  \begin{minipage}{\marginparwidth}
    \centering
    \includegraphics[width=1.0\marginparwidth]{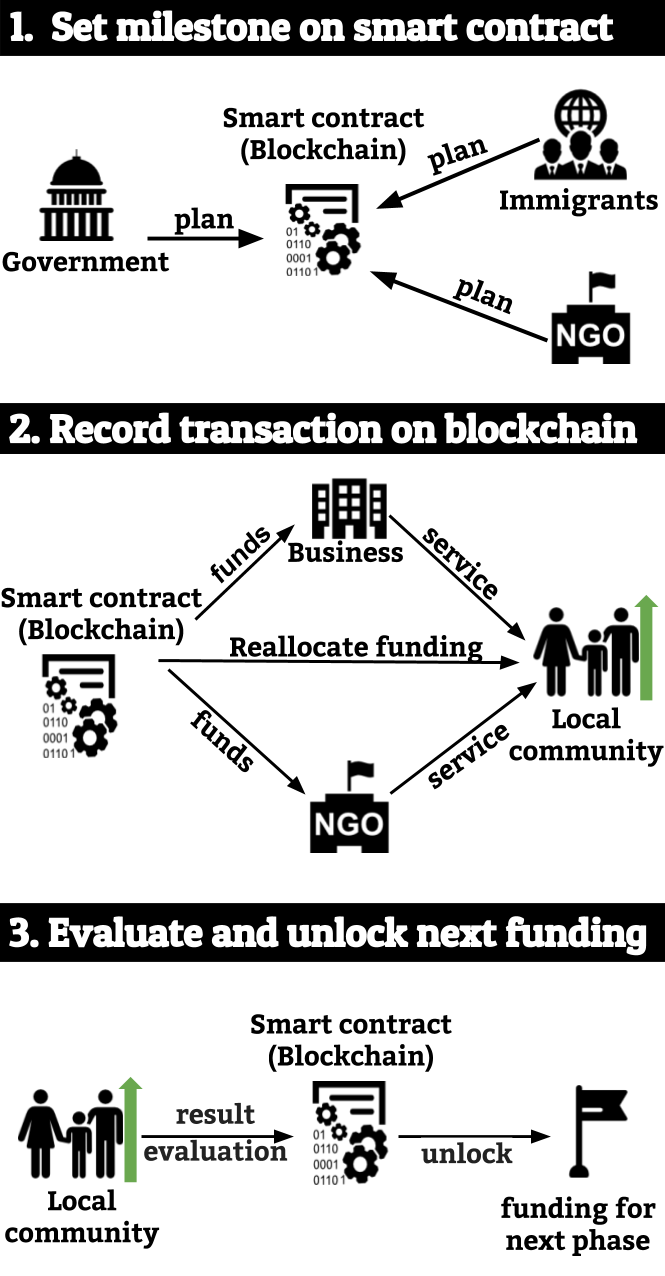}
    \caption{\sys functions on mobile devices and has 3 modules: {\bf  1) Milestone Setting}:  all stakeholders in a decentralized matter propose a plan to help a community, and set the budget of the plan as milestones on a smart contract. {\bf 2) Visualization}: Records all donations and expenditures on the blockchain, and allows public monitoring of fiscal reports data visualizations. {\bf  3) Evaluation}: The system evaluates the state of the project, and decides whether to unlock the next stage of funding. }~\label{fig:overview}
  \end{minipage}
\end{marginfigure}

We introduce \sys, a mobile decentralized collaborative platform for immigrants, governments, and other institutions, such as NGOs, that can rapidly inform individuals about the flow of their money, as well as provide real-time supervision of all transactions. This latter point is especially important as Mexican immigrants generally believe their government is corrupt and will misuse any resources they donate \cite{morris2009political}. To enable real-time supervision and reporting of cash flow, \sys uses smart contracts \cite{buterin2014next}. A Smart contract is a new technology that has been made possible by public blockchains. While normal contracts outline a relationship and enforce the relationship via law, smart contracts enforce the established relationship via cryptographic code. Smart contracts are essentially programs that digitally facilitate, verify, or enforce the negotiation or performance of a contract. We decided to integrate smart contracts into \sys because: (1) every transaction in the blockchain (where the smart contract operates) is traceable and append-only \cite{nakamoto2008bitcoin}; therefore  transactions cannot be falsified and the public can trace the amount and destination of each fund used; (2) smart contracts allow people to set the purpose of each fund and to execute transactions automatically only if the transaction meets the conditions that were set in advance. \sys thus enables the stakeholders of community development projects to designate how they want their funds for a particular project to be used, to ensure that the funds are indeed utilized in the way that was intended, and also enables all parties to keep public track of the execution of the project. Smart contracts also facilitate decentralized collaborations. Figure~\ref{fig:overview} presents an overview of our system.

\section{Discussion and Future Work}

\begin{marginfigure}[0pc]
  \begin{minipage}{\marginparwidth}
    \centering
    \includegraphics[width=1.0\marginparwidth]{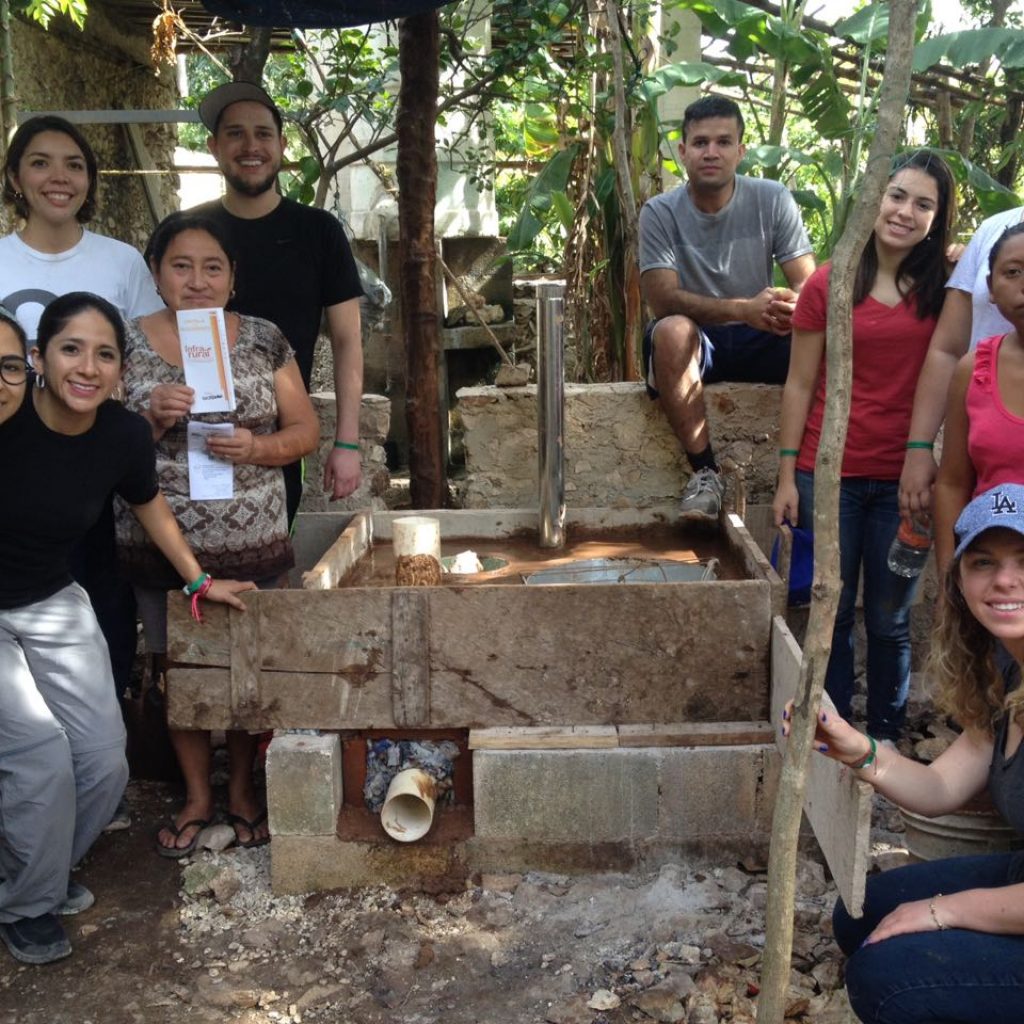}
    
    \caption{ Example of resulting collaborations between immigrants, governments, locals, and social companies (especially the company Infrarural \url{http://infrarural.com/ }). Here the group is collaborating to at scale build cooking infrastructure and help transform Mexican rural communities. \sys empowers immigrants to easily collaborate and trust all players involved and collectively improve their hometowns. } 
    \label{fig:casa}
  \end{minipage}
\end{marginfigure}

Through our analysis, we identified that for Mexican immigrants transparent workflows of how their funds moved was important for trust building. This result matches the recent findings of the Open Government Partnership, which identified that for trust building it was important to offer supervision and accountability \cite{open2018declaration}. We took these findings and designed \sys: a mobile decentralized system that facilitates collaborations between immigrants and governments. \sys  pushes a power balance between governments and immigrants as it allows decentralized collaborations where all stakeholders can establish an execution plan, view all transactions and supervise the execution. In the near future, we will conduct a user study evaluating our system to investigate its effectiveness in building trust. Figure \ref{fig:casa} shows an example of how governments, social companies, and immigrants have started to use our system to collaborate \cite{angel2015participatory}. While researchers have started using blockchain technology to solve existing difficulties in financial and governmental institutions \cite{olnes2017blockchain}, we still lack an understanding of how blockchain could address trust issue between governments and citizens. Our work helps to start investigating this gap.


\subsection{Design Implications for Blockchain Developers.}
One of the features of blockchain is that it is ``trustful'', which means that all the transactions (records) on the blockchain cannot be deleted or falsified. However, this does not guarantee that what is inputted into the blockchain is truthful. It could be that a corrupt official colluded with a company to increase the price of the company's products to keep the extra fees. Designers should consider this problem and think about how to overcome it to design truly trustful technology. 
In \sys we decide to incorporate a milestone-setting phase where immigrants and governments brainstorm their plan and budget, and an evaluation phase where the immigrants can lock the funding when the finances of the project are unclear or the quality of the work is low-grade. This helps citizens to feel more confident about collaborating with the government, as they can  
more easily flag and break corrupt transactions. Blockchain designers should also consider that the value of cryptocurrencies, i.e., the currency units that are used in the blockchain fluctuate greatly and cryptocurrency is also hard to treat as a medium of exchange in the real world. Therefore, it might not be convenient to store the actual funds of the community project on the blockchain. To conquer the fluctuation problem, \sys only used blockchain technology for record keeping rather than for trade. All of the funding for the community development projects are deposited in banks (this is also important given that immigrants trusted banks more). However, this design also creates a new middleman problem, which blockchain technology promised to eliminate \cite{nakamoto2008bitcoin}. We are currently exploring the resolution of this problem through crowdsourcing. Designers also have to analyze people's adoption and use of blockchain technology. For the public, blockchain is still in its initial stages and most do not understand it. In \sys we hid the blockchain aspect of the system and simply presented people with a mobile interface to manipulate. 

\subsection{Design Implications for Civic Platform Developers.}
Prior work had identified that in Latin America there is a general distrust for the government \cite{kuznetsov2006diaspora}. Consequently, transparent technology might not be enough. We believe that to build trust it is important to also push campaigns that present to citizens how corruption is currently being fought. This could help change citizens' mindset that ``corruption is systematic in the country and no technological advancements will transform that reality.'' Such campaigns could e.g., focus on highlighting cases where important public figures were prosecuted for corruption. 
Another aspect for designers to consider is that there might be certain policies or even laws that impede the government from being completely open. Civic platform designers should think about how to effectively communicate these restrictions to end-users as it could also lead to misunderstandings and the belief that the government continues to be corrupt, hindering collaborations. It could also be helpful for civic platform designers to develop mechanisms to help governments be more open about their work dynamics. The lack of such practices can generate unnecessary doubts and affect collaborations. Finally, when developing technology for rural areas, it could also help civic developers to consider theories of alternative development \cite{friedmann1992empowerment}. Alternative development focuses on improving the economic development of an area by targeting the root causes of their problems and giving residents the agency to address the problems, e.g., address that rural citizens might be involved in illicit activities \cite{savage2015participatory} and therefore provide tools to brainstorm and solve that problem. 

{\bf Acknowledgments.} Special thanks to the reviewers, Infrarural.com for their assistance in this research. This work was partially funded by a J. Wayne and Kathy Richards Faculty in Engineering fellowship, FY2018 Research Challenge Grant, research grants from Leidos and Facebook. 


\balance{} 
\small
\bibliographystyle{SIGCHI-Reference-Format}
\bibliography{sample}

\end{document}